\documentclass[%
reprint, onecolumn, superscriptaddress,
showpacs,preprintnumbers,
nofootinbib,
 amsmath,amssymb,
 aps,
pra,
]{revtex4}

\usepackage[all]{xy}
\usepackage{mathrsfs}
\usepackage{graphicx}
\usepackage{epstopdf}
\usepackage{dcolumn}
\usepackage{bm}
\usepackage{color} 
\usepackage{CJK}
\newcommand{\beq}{\begin{equation}}
\newcommand{\eeq}{\end{equation}}
\newcommand{\beqa}{\begin{eqnarray}}
\newcommand{\eeqa}{\end{eqnarray}}

\def\ra{\rangle}
\def\la{\langle}

\usepackage{amsmath,amsfonts,amssymb}
\usepackage{graphicx}
\DeclareGraphicsExtensions{.pdf,.png,.jpg}
\usepackage{bm}
\usepackage{dsfont}

\usepackage{color}

\begin{document}
\title{Fast separation of two trapped ions}
\author{M. Palmero}
\affiliation{Departamento de Qu\'{\i}mica F\'{\i}sica, UPV/EHU, Apdo. 644, 48080 Bilbao, Spain}
\email{mikel.palmero@ehu.eus}
\author{S. Mart\' inez-Garaot}
\affiliation{Departamento de Qu\'{\i}mica F\'{\i}sica, UPV/EHU, Apdo. 644, 48080 Bilbao, Spain}
\author{U. G. Poschinger}
\affiliation{QUANTUM, Institut f\"ur Physik, Universit\"at Mainz, D-55128 Mainz, Germany}
\author{A. Ruschhaupt}
\affiliation{Department of Physics, University College Cork, Cork, Ireland}
\author{J. G. Muga}
\affiliation{Departamento de Qu\'{\i}mica F\'{\i}sica, UPV/EHU, Apdo. 644, 48080 Bilbao, Spain}
\affiliation{Department of Physics, Shanghai University, 200444
Shanghai, People's Republic of China}
%
%
\begin{abstract}
We design fast protocols to separate or recombine two ions in a segmented Paul trap. By inverse engineering 
the time evolution of the trapping potential composed of a harmonic and a quartic term, it is possible to perform these processes in
a few microseconds without final excitation. These times are much shorter than the ones reported so far experimentally. 
The design is based on dynamical invariants and dynamical normal modes. Anharmonicities beyond the harmonic approximation at potential minima 
are taken into account perturbatively.  The stability versus an unknown potential bias is also studied. 
\end{abstract}
\pacs {03.67.Lx, 37.10.Ty, 42.50.Dv}
\maketitle
\section{Introduction}
Trapped cold ions provide a leading platform to implement quantum information
processing.     
Separating ion chains is in the toolkit of basic operations required. (Merging chains is the corresponding reverse operation so we shall only
refer to separation hereafter.)
It has been used to implement two-qubit quantum gates \cite{processor};  
also to purify entangled states \cite{entanglement,entanglement2}, or 
teleport material qubits \cite{teleportation}. 
Moreover,  an architecture for processing information scalable to many ions could be developed based
on shuttling, separating and merging ion crystals in multisegmented traps \cite{kielpinski}.  

Ion-chain separation is known to be a difficult operation \cite{Home}. Experiments have progressed towards lower final excitations and shorter times 
but much room for improvement still remains \cite{Rowe,Bowler,Ruster}. 
Problems identified include anomalous heating, so devising short-time protocols via shortcuts to adiabaticity (STA)  
techniques was proposed as a way-out worth exploring  \cite{Kaufmann}.
STA methods intend to speed up different adiabatic operations \cite{Xi,Review} without inducing final excitations. An example of 
an elementary (fast quasi-adiabatic) STA approach \cite{Xi} was already applied for fast chain splitting in  \cite{Bowler}. 
Here, we design, using a more general and efficient STA approach based on dynamical normal modes \cite{Palmero,expansion}, 
protocols to effectively separate two equal ions, initially in a common electrostatic linear harmonic trap, 
into a final configuration where each ion is in a different well.  The motion is assumed to be effectively one dimensional due to tight radial confinement. 
The external potential for an ion at $q$ is approximated as 
\begin{equation}
V_{ext}=\alpha(t) q^2+\beta(t) q^4,
\end{equation}
which is experimentally realizable 
with state-of-art segmented Paul Traps \cite{Home,NH}.   

Using dynamical normal modes (NM) \cite{Palmero,expansion}  a Hamiltonian will be set  
which is separable in a harmonic approximation around potential minima. 
By means of Lewis-Riesenfeld invariants \cite{LR} we shall design first the approximate dynamics of an unexcited splitting,
taking into account anharmonicities in a perturbative manner,  
and from that inversely find the corresponding 
protocol, i.e. the $\alpha(t)$ and $\beta(t)$ functions. 
%
%
\begin{figure}[b]
\begin{center}
\includegraphics[width=8cm]{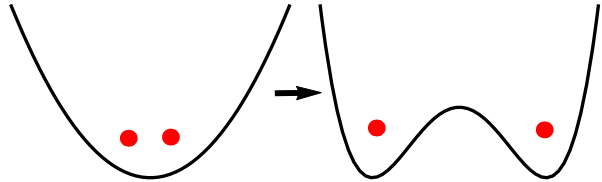}
\caption{\label{scheme}(Color online) 
Scheme of the separation process. At $t=0$ (left), both ions are trapped within the same external harmonic potential. At final time $t_f$ (right), 
the negative harmonic term, and a quartic term build a double well external potential. The aim of the process 
is to set each of the ions in a different well without any residual excitation.}  
\end{center}
\end{figure}
%
%

The Hamiltonian of the system of two ions of mass $m$ and charge $e$ is, in the laboratory frame,  
\begin{eqnarray}
\label{labHamiltonian}
H&=&\frac{p_1^2}{2m}+\frac{p_2^2}{2m}+V,
\nonumber\\
V&=&\alpha (t) (q_1^2+q_2^2)+\beta (t) (q_1^4+q_2^4)+\frac{C_c}{q_1-q_2},
\end{eqnarray}
where $p_1, p_2$ are the momentum operators for both ions, $q_1, q_2$ their position operators, 
and $C_c=\frac{e^2}{4\pi\epsilon_0}$, $\epsilon_0$ being the vacuum permittivity. We use on purpose a c-number notation 
since we shall also consider classical simulations. The context will make clear if $c-$numbers or $q-$numbers are required. 
We suppose that, due to the strong Coulomb repulsion, $q_1>q_2$. 
By minimizing the potential part of the Hamiltonian $V$, we find for the equilibrium distance 
between the two ions, $d(t)$, the quintic equation \cite{Home}
\begin{equation}
\label{beta}
\beta(t)d^5(t)+2\alpha(t)d^3(t)-2C_c=0, 
\end{equation}
%
which will be quite useful for inverse engineering the ion-chain splitting, even without an explicit solution for $d(t)$.   
At $t=0$ a single external well is assumed, $\beta (0)=0$ and $\alpha(0)>0$, whereas in the final double-well configuration $\beta (t_f)>0, \alpha(t_f)<0$. 
At some intermediate time $t_i$ the potential becomes  purely quartic ($\alpha (t_i)=0$). 
Our aim is to design the functions $\alpha(t)$ and $\beta(t)$ so that each of the ions ends up in a different external well as shown in Fig. \ref{scheme},
in times as short as possible, and without any final excitation.   
\section{Dynamical Normal Modes}
To define dynamical NM coordinates, we calculate first at equilibrium (the point $\{q_1^{(0)}, q_2^{(0)}\}$ in configuration 
space where the potential is a minimum, $\partial V/\partial q_1=\partial V/\partial q_2=0$) 
the matrix $V_{ij}=\frac{1}{m}\frac{\partial ^2V}{\partial q_i\partial q_j}\big|_{_{\rm{eq}}}$. 
The equilibrium positions are $q_1^{(0)}=\frac{d (t)}{2}$, $q_2^{(0)}=-\frac{d(t)}{2}$, and the matrix takes the form
\begin{equation}
\label{potential}
V_{ij}=\frac{1}{m}
\begin{pmatrix}
2\alpha +12\beta\frac{d^2}{4}+\frac{2C_c}{d^3}&-\frac{2C_c}{d^3}
\\
\frac{-2C_c}{d^3}&2\alpha +12\beta\frac{d^2}{4}+\frac{2C_c}{d^3}
\end{pmatrix}.
\end{equation}
The eigenvalues are
\begin{eqnarray}
\label{lambdaeq}
\lambda_-&=&\frac{1}{m}(2\alpha +3\beta d^2),
\nonumber\\
\lambda_+&=&\frac{1}{m}\left(2\alpha +3\beta d^2+\frac{4C_c}{d^3}\right),
\end{eqnarray}
which define the NM frequencies as $\Omega_\pm=\sqrt{\lambda_\pm}$ corresponding to center-of-mass ($-$) and relative (stretch) 
motions (+).  
These relations, with Eq. (\ref{beta}) written as 
\begin{equation}
\label{beta2}
\beta(t)=\frac{2C_c}{d^5(t)}-\frac{2\alpha(t)}{d^2(t)},
\end{equation}
allow us to write $\alpha(t)$ and $d(t)$ as functions of the NM frequencies,  
\begin{eqnarray}
\label{alpha}
\alpha(t)&=&\frac{1}{8}m\left(3\Omega_+^2-5\Omega_-^2\right),
\\
%
%
\label{distance}
d(t)&=&\sqrt[3]{\frac{4C_c}{m\left(\Omega_+^2-\Omega_-^2\right)}}.
\end{eqnarray}
Substituting these expressions into Eq. (\ref{beta2}), $\beta(t)$ may also be written in terms of NM
frequencies.   
%
%

The normalized eigenvectors are
\begin{eqnarray}
\label{vectoreq}
v_-&=&\frac{1}{\sqrt{2}}\binom{1}{1},
\nonumber\\
v_+&=&\frac{1}{\sqrt{2}}\binom{1}{-1}, 
\end{eqnarray}
which we denote as  $v_\pm=\binom{a_\pm}{b_\pm}$. 
The (mass weighted) dynamical NM coordinates are defined in terms of the laboratory coordinates as 
\begin{equation}
{\sf q}_\pm =a_\pm \sqrt{m}(q_1-q_1^{(0)})+b_\pm\sqrt{m}(q_2-q_2^{(0)}).
\end{equation}
The unitary transformation of coordinates is
\begin{equation}
\label{unitary}
U= \int d{\sf q}_+d{\sf q}_-dq_1dq_2 |{\sf q}_+,{\sf q}_-\rangle\langle {\sf q}_+,{\sf q}_-| q_1,q_2\rangle \langle q_1,q_2|,
\end{equation}
where  $\langle {\sf q}_+,{\sf q}_-|q_1,q_2\rangle =\delta [q_1-q_1({\sf q}_+,{\sf q}_-)]\delta [q_2-q_2({\sf q}_+,{\sf q}_-)]$.
The Hamiltonian in the dynamical equation for $|\psi'\ra=U|\psi\ra$, where $|\psi\ra$ is the lab-frame time-dependent wave function evolving with $H$, 
is given by
\begin{eqnarray}
\label{Hprime}
H'&=&UHU^\dagger -i\hbar U(\partial_t U^\dagger)=\nonumber\\
&=&\frac{{\sf p}_+^2}{2}+\frac{1}{2}\Omega_+^2{\sf q}_+^2+\frac{\dot{d}}{\sqrt{2}}\sqrt{m}{\sf p}_+
\nonumber\\
&+&\frac{{\sf p}_-^2}{2}+\frac{1}{2}\Omega_-^2{\sf q}_-^2,
\end{eqnarray}
plus qubic and higher order terms in the potential that we neglect by now (they will be considered in Sec. IV below).  
Similarly to \cite{Palmero,expansion}, we apply a further unitary transformation  $\mathcal{U}=e^{-i\sqrt{m}\dot{d}{\sf q}_+/(\sqrt{2}\hbar)}$, 
to write down an effective  Hamiltonian for $|\psi''\ra=\mathcal{U}|\psi'\ra$ with the form of two independent harmonic oscillators in NM space, $H''=\mathcal{U}H'\mathcal{U}^\dagger -i\hbar \mathcal{U}(\partial_t\mathcal{U}^\dagger)$,
\begin{eqnarray}
\label{HamNM}
H''&=&\frac{{\sf p}_+^2}{2}+\frac{1}{2}\Omega_+^2\left({\sf q}_++\frac{\sqrt{m}\ddot{d}}{\sqrt{2}\Omega_+^2}\right)^2
\nonumber\\
&+&\frac{{\sf p}_-^2}{2}+\frac{1}{2}\Omega_-^2{\sf q}_-^2=H''_++H''_- .
\end{eqnarray}
These oscillators have  dynamical invariants of the form \cite{LR}
\begin{eqnarray}
\label{Invariant}
I_\pm&=&\frac{1}{2}[\rho_\pm ({\sf p}_\pm-\dot{x}_\pm)-\dot{\rho}_\pm({\sf q}_\pm-x_\pm)]^2
\nonumber\\
&+&\frac{1}{2}\Omega_{0\pm}^2\left(\frac{{\sf q}_\pm-x_\pm}{\rho_\pm}\right)^2, 
\end{eqnarray}
where the auxiliary functions $\rho_\pm$  and $x_+$  
satisfy 
\begin{eqnarray}
\label{auxiliaryerm}
\ddot{\rho}_\pm+\Omega_\pm^2\rho_\pm=\frac{\Omega_{0\pm}^2}{\rho_\pm^3},
\\
\label{auxiliarynew}
\ddot{x}_++\Omega_+^2x_+=-\sqrt{\frac{m}{2}}\ddot{d},
\end{eqnarray}
with $\Omega_{0\pm}=\Omega_\pm (0)$, and, due to symmetry, $x_-=0$. 
 
The physical meaning of the auxiliary functions 
may be grasped from  the solutions of the time-dependent Schr\"odinger
equations (for each NM Hamiltonian $H''_\pm$ in Eq. (\ref{HamNM})) proportional to the invariant eigenvectors \cite{Erik}. 
They form a complete basis for the space spanned by each Hamiltonian $H''_\pm$
and take the form
\begin{equation}
\label{expandingmodes}
\la {\sf q}_\pm|\psi'' _{n\pm}(t)\rangle=e^{\frac{i}{\hbar} \left[\frac{\dot{\rho}_\pm{\sf q}_\pm^2}{2\rho_\pm}+(\dot{x}_\pm\rho_\pm-x_\pm\dot{\rho}_\pm)
\frac{{\sf q}_\pm}{\rho_\pm}\right]}
\frac{\Phi_n(\sigma_\pm)}{\rho_\pm^{1/2}},
\end{equation}
where $\sigma_\pm=\frac{{\sf q}_\pm-x_\pm}{\rho_\pm}$, and $\Phi_n(\sigma_\pm)$ are the eigenfunctions of the static harmonic oscillator at time $t=0$. 
Thus $\rho_\pm$ are scaling factors proportional to the state ``width'' in NM coordinates,  whereas the $x_\pm$ are the dynamical-mode centers in the space of NM coordinates.  
Within  the harmonic approximation there are dynamical states of the factorized form 
$|\psi''(t)\rangle=|\psi''_+(t)\rangle |\psi''_-(t)\rangle$ for the ion chain dynamics, 
where the  NM wave functions $|\psi''_\pm(t)\ra$
evolve independently with $H''_\pm$.
They may be written as combinations of the form $|\psi''_{\pm}(t)\ra=\sum_n C_{n\pm} |\psi''_{n\pm}(t)\ra$,
with constant amplitudes $C_{n\pm}$. 
The average energies of  the $n$-th basis states for the two NM are 
$E''_{n\pm}=\langle \psi''_{n\pm} |H''_\pm|\psi''_{n\pm} \rangle$,
\begin{eqnarray}
\label{energy}
E''_{n-}&=&\frac{(2n+1)\hbar}{4\Omega_{0-}}\left(\dot{\rho}_-^2+\Omega_-^2\rho_-^2+\frac{\Omega_{0-}^2}{\rho_-^2}\right),
\nonumber\\
E''_{n+}&=&\frac{(2n+1)\hbar}{4\Omega_{0+}}\left(\dot{\rho}_+^2+\Omega_+^2\rho_+^2+\frac{\Omega_{0+}^2}{\rho_+^2}\right)
\nonumber\\
&+&\frac{1}{2}\dot{x}_+^2+\frac{1}{2}\Omega_+^2\left(x_+-\frac{\sqrt{m}\ddot{d}}{\sqrt{2}\Omega_+^2}\right)^2.
\end{eqnarray}
\section{Design of the Control Parameters}
Once the Hamiltonian and  Lewis-Riesenfeld  invariants are defined, we proceed to apply the invariant based inverse engineering technique and design shortcuts to adiabaticity. The results for the simple harmonic oscillator in \cite{Xi} serve as a reference but have to be extended
here since the two modes are not really independent from the perspective of the inverse problem. This is because 
a unique protocol, i.e., a single set of $\alpha(t)$  and $\beta(t)$ functions has to be designed.   

%

We first set the initial and target values for the  control parameters $\alpha(t)$ and $\beta(t)$. At time $t=0$, the external trap -for a single ion- is purely harmonic, with (angular) frequency $\omega_0$. From Eq. (\ref{lambdaeq}), we find  that $\Omega_-(0)=\omega_0$ and $\Omega_+(0)=\sqrt{3}\omega_0$.  The equilibrium distance is $d(0)=\sqrt[3]{\frac{2C_c}{m\omega_0^2}}$. For the final time, we 
set a tenfold expansion of the equilibrium distance, $d(t_f)=10d(0)$, and $\Omega_-(t_f)=\omega_0$. This also implies $\Omega_+(t_f)=\sqrt{1.002}\, \omega_0\approx\Omega_-(t_f)$, \emph{i.e.},    
the final frequencies of both NM are essentially equal, the Coulomb interaction is negligible, 
and the ions can be considered to oscillate in 
independent traps. 

%
\begin{figure*}[t!]
\begin{center}
\includegraphics[height=2.8cm]{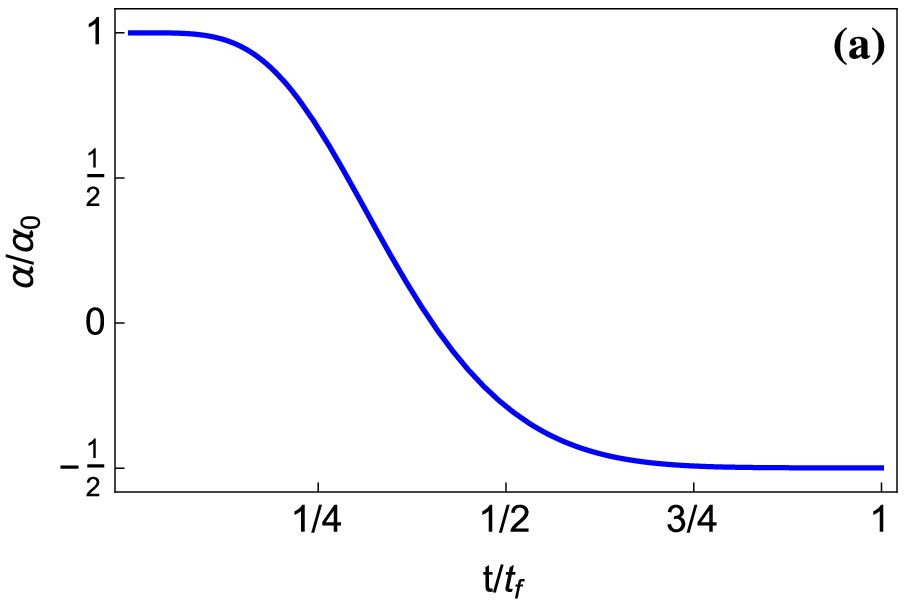}
\includegraphics[height=2.8cm]{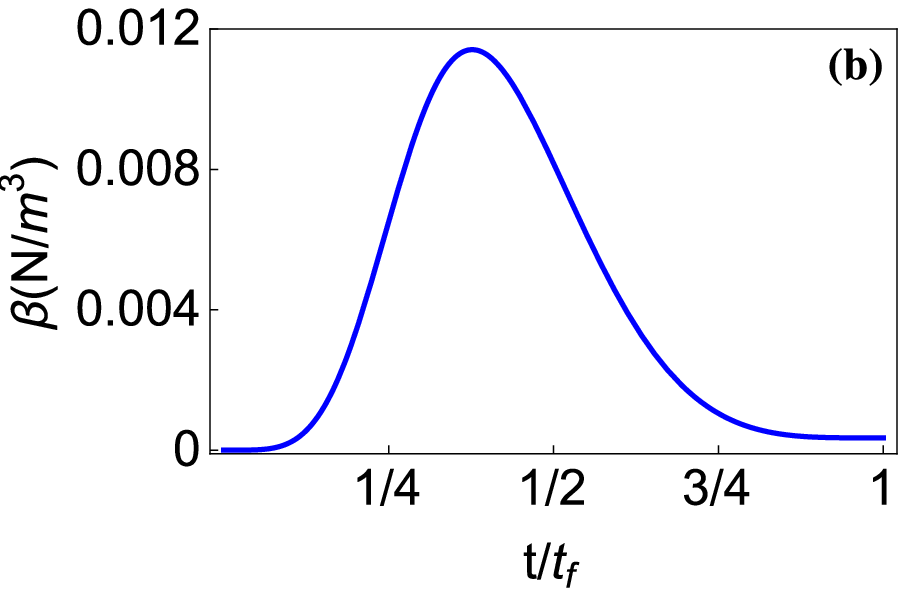}
\includegraphics[height=2.8cm]{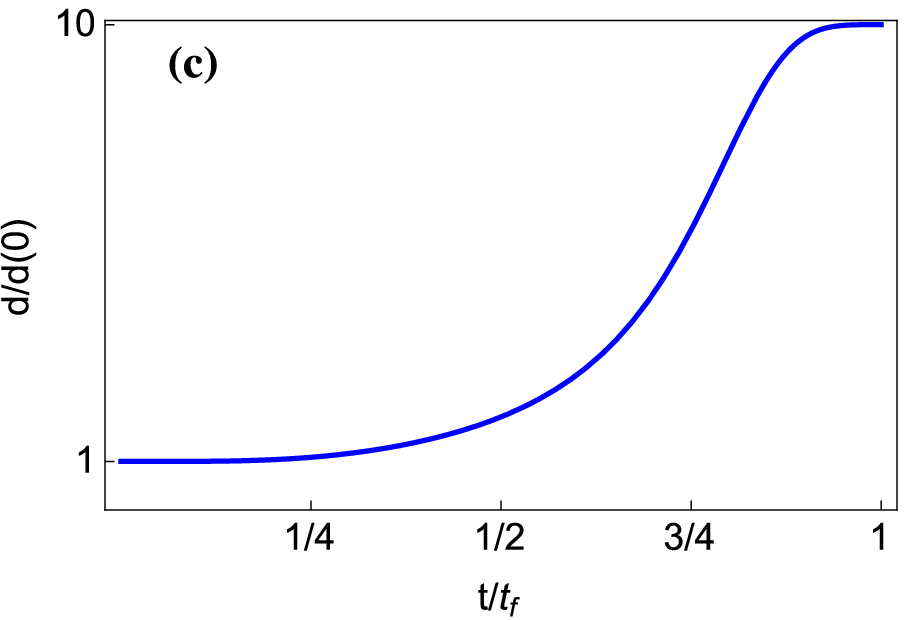}
\includegraphics[height=2.8cm]{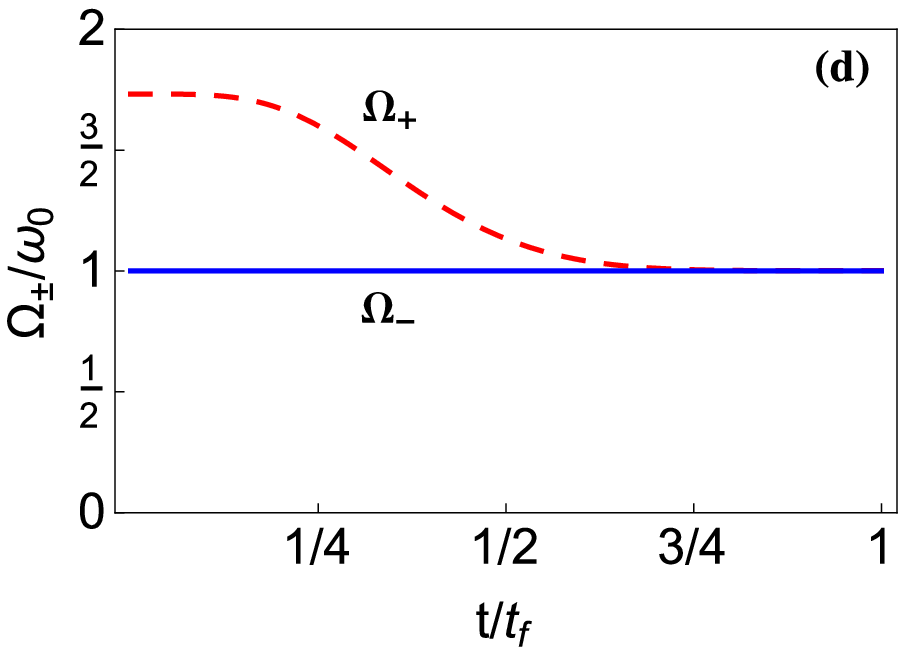}
\caption{\label{parameters}(Color online) 
Evolution of a) $\alpha(t)$;   b) $\beta (t)$; and c) d(t). 
In d) the NM frequencies, solid line for the `-' and dashed line for the `+' are depicted.  
Two $^9$Be$^+$ ions were separated in the simulation, with $\omega_0/(2\pi)=2$ MHz, $t_f=5.2 $ $\mu$s, $\alpha_0=m\omega_0^2/2$, and $d(0)=5.80$ $\mu$m.}  
\end{center}
\end{figure*}
%
The inverse problem is somewhat similar to the expansion of a trap with two equal ions in Ref. \cite{expansion},
but complicated by the richer structure of the external potential.  
The Hamiltonian (\ref{HamNM}) and the invariant (\ref{Invariant}) must commute at both boundary times $[H(t_b),I(t_b)]=0$, 
\begin{equation}
t_b=0,t_f,
\end{equation}
to  drive initial levels into final levels via a one-to-one mapping. 
This is achieved by applying appropriate boundary conditions (BC) to the auxiliary functions $\rho_\pm$, $x_\pm$ and
their derivatives, 
%
%
%
\begin{eqnarray}
\rho_\pm(0)&=&1,\;\; \rho_\pm(t_f)=\gamma_\pm, 
\label{rho1}
\\
\dot{\rho}_\pm(t_b)&=&\ddot{\rho}_\pm(t_b)=0,
\label{rho2}
\\
\label{xbc}
x_+(t_b)&=&\dot{x}_+(t_b)=\ddot{x}_+(t_b)=0,
\end{eqnarray}
where $\gamma_\pm=\sqrt{\frac{\Omega_\pm(0)}{\Omega_\pm(t_f)}}$. Let us recall that $x_-=0$ for all times 
so this parameter does not have to be 
considered further.

Inserting the BC for $x_+(t_b)$ and $\ddot{x}_+(t_b)$ in Eq. (\ref{auxiliarynew}) we find that  $\ddot{d}(t_b)=0$. Additionally, $\dot{d}(t_b)=0$ is to be imposed so that ${\mathcal{U}}(t_b)=1$.
According to Eq. (\ref{distance}), $\dot{d}(t_b)=\ddot{d}(t_b)=0$ by imposing $\dot{\Omega}_\pm(t_b)=\ddot{\Omega}_\pm(t_b)=0$. 
With $\dot{d}(t_b)=\ddot{d}(t_b)=0$ the Hamiltonians and wave functions coincide at the boundaries, $H'(t_b)=H''(t_b)$, 
$|\psi'(t_b)\ra=|\psi''(t_b)\ra$, which simplifies the calculation of the excitation energy.      

From Eq. (\ref{auxiliaryerm}), the NM frequencies may be written as 
\begin{equation}
\label{omega}
\Omega_\pm=\sqrt{\frac{\Omega_{0\pm}^2}{\rho_\pm^4}-\frac{\ddot{\rho}_\pm}{\rho_\pm}}.
\end{equation}
Thus the BC $\dot{\Omega}_\pm(t_b)=\ddot{\Omega}_\pm(t_b)=0$ are satisfied by imposing on   
the auxiliary functions the additional BC 
\begin{equation}
\label{rho3}
\dddot{\rho}_\pm(t_b)=\ddddot{\rho}_\pm(t_b)=0.
\end{equation}
%

We may now design ansatzes for the auxiliary functions $\rho_\pm$ that satisfy the ten BC in Eqs. (\ref{rho1},\ref{rho2},\ref{rho3}), plus the  BC for $x_+(t_b)$ and $\dot{x}_+(t_b)$  in Eq. (\ref{xbc}) (since $\ddot{d}(t_b)=0$, $\ddot{x}_+(t_b)=0$ is then automatically satisfied, see Eq. (\ref{auxiliarynew})).  
Finally, from the NM frequencies given by  
Eq. (\ref{omega}) we can inverse engineer the control parameters $\alpha(t)$ and $\beta(t)$ from Eqs. (\ref{alpha},\ref{distance}) and (\ref{beta2}). 

A simple choice for $\rho_-(t)$ is a polynomial ansatz of 9th order $\rho_{-}=\sum_{i=0}^9b_is^i$, where 
$s={t}/{t_f}$. Substituting this form in the 
ten BC in Eqs. (\ref{rho1},\ref{rho2},\ref{rho3}),  we finally get
\begin{eqnarray}
\label{ansatz1}
\rho_-&=&126(\gamma_--1)s^5-420(\gamma_--1)s^6+540(\gamma_--1)s^7
\nonumber\\
&-&315(\gamma_--1)s^8+70(\gamma_--1)s^9+1.
\end{eqnarray}
For $\rho_+$ we will use an 11-th order polynomial $\rho_+=\sum_{n=0}^{11}a_ns^n$ to satisfy 
as well  
$x_+(t_b)=\dot{x}_+(t_b)=0$. The parameters $a_{0-9}$ are fixed so that  the 10 BC for $\rho_+$ are fulfilled (see the Appendix), whereas $a_{10}$, $a_{11}$ are left free, and will be numerically determined by a shooting program \cite{Shooting} (`fminsearch'  in MATLAB which uses the Nelder-Mead simplex method for optimization), so that the remaining BC for $x_+(t_b)$ and $\dot{x}_+(t_b)$ are also satisfied.
Specifically, for each pair $\{a_{10}, a_{11}\}$, $\Omega_\pm(t)$ and $d(t)$ are determined 
from Eqs. (\ref{auxiliaryerm}) and (\ref{distance}), 
to solve Eq. (\ref{auxiliarynew}) for $x_+(t)$ with initial conditions $x_+(0)=\dot{x}_+(0)=0$. The free constants are changed until  $x_+(t_f)=0$ and $\dot{x}_+(t_f)=0\,$ are satisfied. Numerically a convenient way to find the solution is to minimize 
the energy $E''_{n+}(t_f)$ in Eq. (\ref{energy}). 
 
%

\begin{figure}[t!]
\begin{center}
\includegraphics[width=5.5cm]{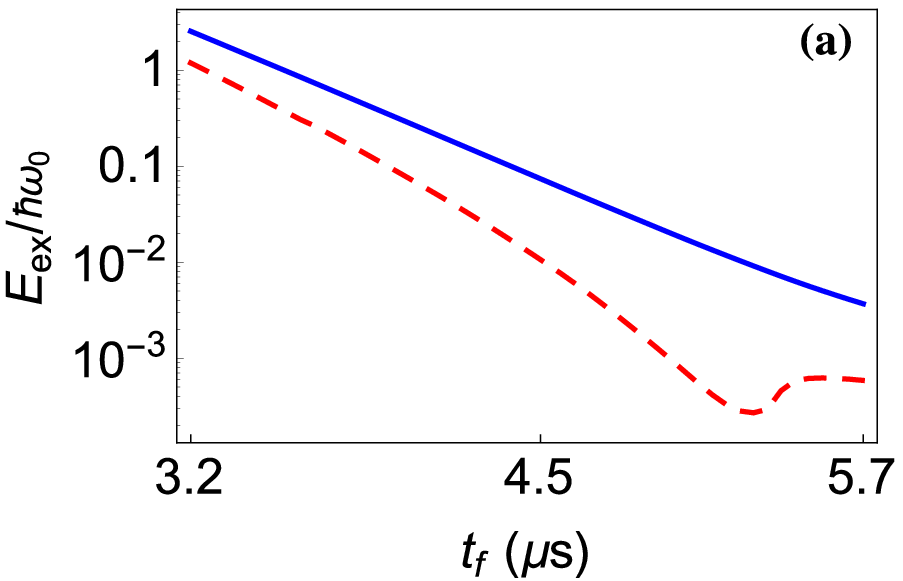}
\includegraphics[width=5.5cm]{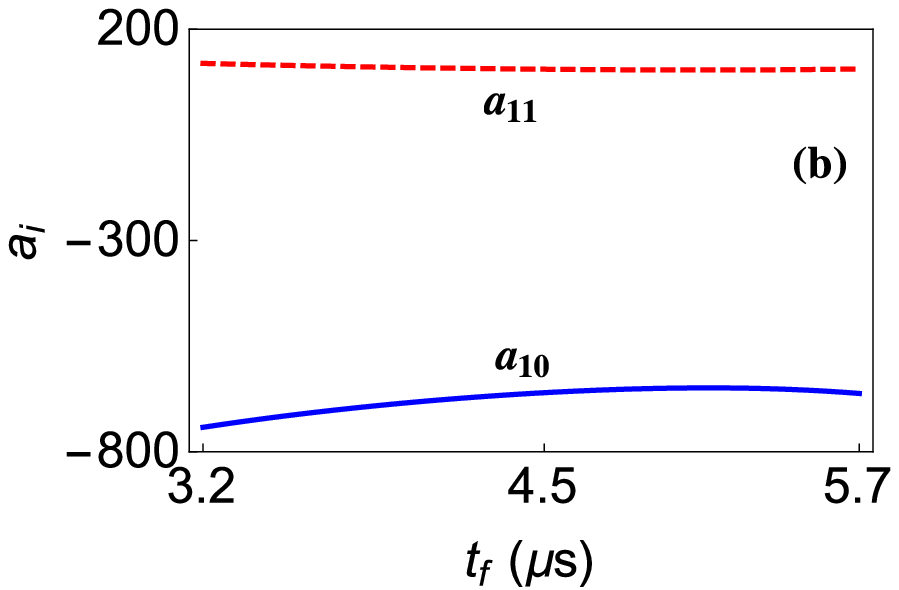}
\includegraphics[width=5.5cm]{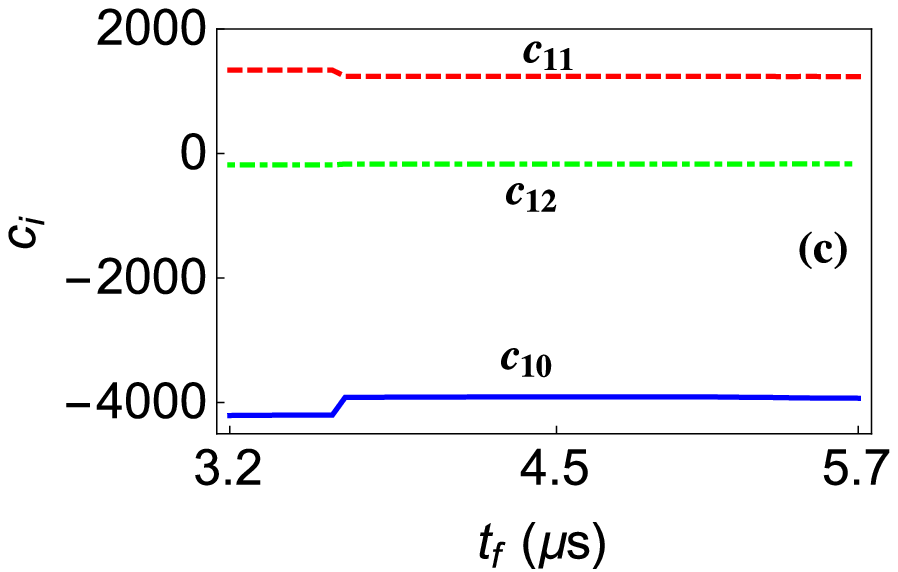}
\caption{\label{excitation}(Color online) 
a) Final excitation energy vs. final time using the inverse engineering design of Sec. III (solid blue),
and the design that takes into account anharmonicities in Sec. IV (dashed red).
b) Values of the free parameters $a_{10}$ (solid blue) and $a_{11}$ (dashed red) that minimize the excitation energy for the 11th order polynomial (\ref{ansatz2}). c) Parameters $c_{10}$ (solid blue), $c_{11}$ (dashed red) and $c_{12}$ (dash-dotted green) that  minimize the excitation energy for the 12th order polynomial (\ref{ansatz3}). 
Two $^9$Be$^+$ ions where split, with $\omega_0/(2\pi)=2$ MHz.}  
\end{center}
\end{figure}

Fig. \ref{parameters} (a) and (b) depict the control parameters $\alpha(t)$ and $\beta(t)$ found with this method, using Eqs. 
(\ref{alpha}) and (\ref{beta2}), for some value of $t_f$ and $\omega_0$,
see the caption,  while Fig. \ref{parameters} (c) represents the equilibrium distance between ions as a function of time (\ref{distance}), and Fig. \ref{parameters} (d)  the NM frequencies. In Fig. \ref{excitation} (a) the excitation energy is shown, versus final time, for optimized parameters    
given in Fig. \ref{excitation} (b). The initial state is the ground state of the two ions.  It is calculated with an ``imaginary time evolution''. 
This is a variation of any numerical time evolution method (here we used a variation of the split-operator method), where instead of the real time one defines imaginary time in the evolution operator. By letting an ansatz wave function evolve for the static initial potential, it will eventually converge to the ground state of the system.
The excitation energy is  $E_{ex}=E(t_f)-E_0(t_f)$, where $E(t_f)$
is the final energy, calculated  in the lab frame, and $E_0(t_f)$ is the final ground-state energy. The wave function evolution is calculated using the ``Split-Operator Method'' with the full Hamiltonian (\ref{labHamiltonian}).  
If the harmonic approximation were exact, there would not be any excitation with this STA method, 
$E(t_f)=E''_{0+}(t_f)+E''_{0-}(t_f)=E_0(t_f)$, see Eq. (\ref{energy}). The actual result is perturbed by the anharmonicities and NM couplings.  The final ground state is also calculated with an ``imaginary time evolution''. 
The corresponding final ground state energy is essentially two times a harmonic oscillator ground state energy
plus the (negligible) Coulomb repulsion at  distance $d(t_f)$. For the final times of all the examples, as it was noted in previous works \cite{Kaufmann,Palmero,Palmero2,expansion}, 
classical simulations (solving Hamilton's equations from the equilibrium configuration instead of Schr\"odinger's equation) 
give indistinguishable results in the scale of  Figure \ref{excitation} (a).  

The excitation energy in Figure \ref{excitation}(a) (solid line) increases at short times
since the harmonic NM approximation fails \cite{Palmero,expansion}. However, it goes down rapidly below one excitation quantum at times 
which are still rather small  compared to experimental values used so far \cite{Bowler,Ruster}. 
In the following section we shall apply a perturbative technique to 
minimize the excitation further. 
%
%
%
%
%
%
\section{Beyond the harmonic approximation}
An improvement of the protocol is to consider the perturbation of the higher order terms  
neglected in the Hamiltonian (\ref{HamNM}). These  ``anharmonicities''  \cite{Morigi} are cubic and higher 
order terms in the Taylor expansion of the Coulomb term ${C_c}/({q_1-q_2})$,
\begin{eqnarray}
\delta V&=&\sum_{j=3}^\infty \delta V^{(j)}
\nonumber
\\
&=&\sum_{j=3}^\infty\! \frac{(-1)^jC_c}{d^{j+1}}\!\left[\!\left(q_2-q_2^{(0)}\right)\!-\!\left(q_1-q_1^{(0)}\right)\!\right]^j. 
\end{eqnarray}
In NM coordinates the terms take the simple form
\begin{equation}
\label{pertubation}
\delta V^{(j)}=(-1)^{j+1}\frac{C_c}{d^{j+1}}\left(\sqrt{\frac{2}{m}}{\sf q}_+\right)^j,
\end{equation}
which may be regarded as a perturbation to be added to $H''_+$ in Eq. (\ref{HamNM}).  
(The perturbation does not couple the center-of-mass and relative subspaces.) 
To first order, the excess energy due to these perturbative terms at final time is given by
\begin{equation}
\label{perturbationenergy}
\delta E_{n+}^{(j)}=\langle \psi'' _{n+}(t_f)|\delta V^{(j)}|\psi''_{n+}(t_f)\rangle,
\end{equation}
where the $|\psi''_{n+}\ra$ are the unperturbed states in Eq. (\ref{expandingmodes}). 
Inverse engineering the splitting process may now be carried out by considering a 12th order polynomial for $\rho_+$ (see (\ref{ansatz3})),
with three free parameters so as to fix the BC for $x_+$ and also minimize  
the excitation energy. In practice we use MATLAB's `fminsearch' function for the shooting to minimize  $E_{0+}(t_f)+\delta E_{0+}^{(3)}$ as no significant improvement occurs by including higher order terms. 
Fig. \ref{excitation} (a) compares the performance of such a protocol with the simpler one with
the 11th-order polynomial (\ref{ansatz2}).  Fig. \ref{excitation} (c) gives the values of optimized parameters at different final times.
\section{Discussion}
A large quartic potential is desirable to control the excitations produced at the point where the harmonic term changes its sign \cite{Kaufmann}. At this point, the harmonic potential switches from confining to repulsive, which reduces the control of the system and potentially increases diabaticities and heating. 
In the inverse approach proposed here there is no special design of the protocol at this point, but the method naturally seeks high quartic confinements there.
In Fig. \ref{parameters} (b) $\beta$ reaches its maximum value right at the time where $\alpha$ changes sign (see Fig. \ref{parameters} (a)). 
However, the maximum value that $\beta$ can reach will typically be limited in a Paul trap \cite{Home}.
In Table \ref{table} we summarize the different maximal values of $\beta$  and critical times (final times at which  excitations below 0.1 quanta are reached) for different values of $\omega_0$ using  
the 11-th order polynomial (\ref{ansatz2}) for $\rho_+$.
\begin{table}[t]
\centering
\begin{tabular}{ c | c | c }
$\omega_0$ (MHz) &  $\beta_{max}$ ($10^{-3}$N/m$^3$) & $t_{crit}$ ($\mu$s) \\ \hline
3 & 44.2 & 2.9 \\ \hline
2 & 11.4 & 4.4 \\ \hline
1.2 & 2.082 & 7.4 \\ \hline
0.8 & 0.539 & 11.2 \\ \hline
\end{tabular}
\caption{Maximum values of $\beta$, and critical times (final times at which  excitations below 0.1 quanta are reached) for different values of $\omega_0$. The calculations were performed with the 11-th order polynomial for $\rho_+$.}
\label{table} 
\end{table}
%
%
The maximum $\beta$ decreases with $t_f$, such that the shortest possible $t_f$ at a given maximum tolerable excitation energy  is limited by the achievable $\beta$. The trap used in Ref. \cite{Ruster} yields a maximum $\beta$ of about $10^{-4}$ N/m$^{3}$, at $\pm 10$ V steering range. In a recent experiment reported in \cite{Wilson}, where although the purpose was not ion separation a double well potential was produced, the value used was $\beta\approx 5\times 10^{-3}$ N/m$^3$.
The numbers reported in the Table  are thus within reach, as the $\beta$ coefficients scale with the inverse 4th power of the overall trap dimension, 
and technological improvements on arbitrary waveform generators may allow for operation at an increased voltage range. 

\begin{figure}[t!]
\begin{center}
\includegraphics[width=8cm]{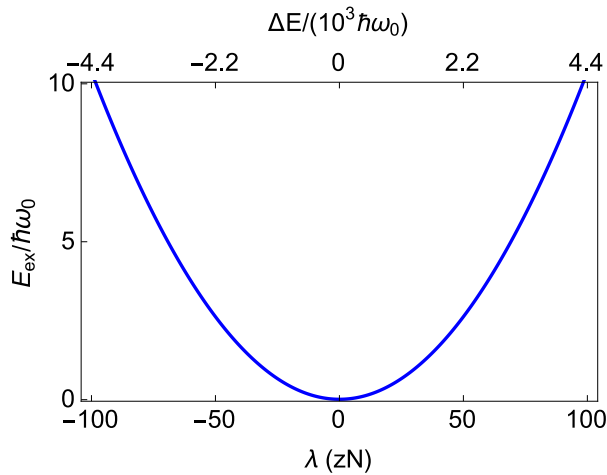}
\caption{\label{bias}(Color online) 
Excitation energy vs. different tilt values of the external potential in terms of the energy difference between both wells (upper axis) and values of the $\lambda$ parameter (lower axis), when using the 11-th order polynomial in the evolution. Same parameters as in Fig. \ref{parameters}.}  
\end{center}
\end{figure}

Another potential limitation the method could encounter in the laboratory is due to biases (a linear slope) 
in the trapping potential, $V_{ext}=\alpha q^2+\beta q^4+\lambda q$, with $\lambda$ constant and unknown \cite{Kaufmann}. 
%
Fig. \ref{bias} represents the excitation energy versus the energy difference between the two final minima of the external potential, $\Delta E$ (also vs. $\lambda$). 
To calculate the results, $\alpha(t)$ and $\beta(t)$ are designed as if $\lambda=0$,
but the dynamics is carried out with a non-zero $\lambda$, in particular the initial state is the actual ground state, including 
the perturbation. Note that $\Delta E$ should be more than a thousand vibrational quanta to excite the final energy by one quantum.   
In Ref. \cite{Ruster} an energy increase of ten phonons at about 150 zN and 80 $\mu$s separation time was reported, so the STA ramps
definitely improve the robustness against bias.  
%

%
%
%
\begin{figure}[t!]
\begin{center}
\includegraphics[width=8cm]{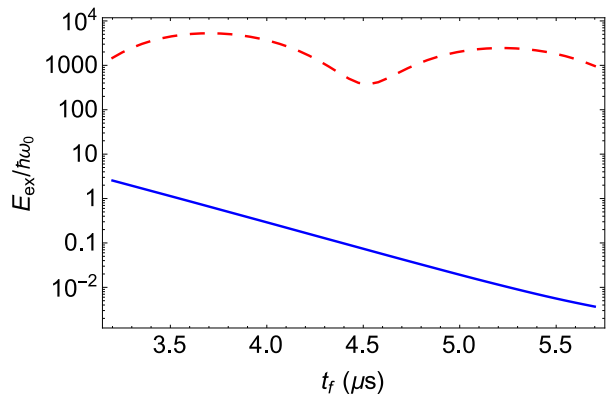}
\caption{\label{comparison}(Color online) 
Excitation energy vs. final time comparing the 11-th order polynomial (solid blue) and a non optimized trajectory experimentally used in \cite{Ruster} (dashed red) in the evolution. Same parameters as in Fig. \ref{parameters}.}  
\end{center}
\end{figure}
%
Further experimental limitations may be due to random fluctuations in the potential parameters, 
or higher order terms in the external potential. We leave these important issues for a separate study 
but note that the structure of the STA techniques used here is well adapted to deal with noise or 
perturbations
\cite{Andreas,Xiao1,Xiao2}.   

Finally, we compare in Fig. \ref{comparison} the performance of the protocols based on the polynomials (\ref{ansatz1}) and (\ref{ansatz2}) with a simple non-optimized protocol based on those experimentally used in \cite{Ruster}. There, the equilibrium distance $d$ is first designed as $d(t)=d(0)+[d(t_f)-d(0)]s^2\sin(s\pi/2)$, where $s=t/t_f$. 
From the family of possible potential ramps consistent with this function, we chose a polynomial that drives $\alpha$ from $\alpha(0)=\alpha_0$ to $\alpha(t_f)=-\alpha_0/2$ (as in Fig. \ref{parameters}) and whose first derivatives are 0 at both boundary times. $\beta$ is given by Eq. (\ref{beta}). For the times analysed in Fig. \ref{comparison}, the method based on Eqs. (\ref{ansatz1}) and (\ref{ansatz2}) clearly outperforms the non-optimized ramp. To get excitations below the single motional excitation  quantum with the non-optimized  protocol, final times as long as $t_f\sim 80\, \mu$s would be needed, which is in line with current experiments. 

We conclude that the method presented here, could bring a clear improvement with respect to the best results experimentally reported so far \cite{Bowler,Ruster}. The parameters required are realistic in current trapped ions laboratories. The simulations show that, under ideal conditions, the separation of two ions could be performed in a few oscillation periods, at times similar to those required for other operations as transport \cite{Palmero} or expansions \cite{expansion}, also studied with STA. 
\section*{Acknowledgements}
This work has been possible thanks to the close collaboration with 
three ion-trap groups at Boulder, Mainz and Zurich. We acknowledge in particular discussions with 
Didi Leibfried, Ludwig de Clercq, Joseba Alonso, and Jonathan Home
that provided essential elements to develop the approach.   
It was supported by the Basque Country Government (Grant No. IT472-10), 
Ministerio de Econom\'\i a y Competitividad (Grant No. FIS2012-36673-C03-01), the program UFI 11/55 of UPV/EHU.
M.P. and S. M.-G. acknowledge fellowships by UPV/EHU. 
This research was funded by the Office of the Director of National Intelligence (ODNI), Intelligence Advanced
Research Projects Activity (IARPA), through the Army Research Office grant W911NF-10-1-0284.
All statements of fact, opinion or conclusions contained herein are those of the authors
and should not be construed as representing the official views or policies of IARPA, the ODNI,
or the US Government.
\begin{appendix}
\section{Ansatz for $\rho_+$}
The ansatz for $\rho_+$ that 
satisfies 
the BC $\rho_+(0)=1$, $\rho_+(t_f)=\gamma_+$, $\dot{\rho}_+(t_b)=\ddot{\rho}_+(t_b)=\dddot{\rho}_+(t_b)=\ddddot{\rho}_+(t_b)=0$
with two free parameters takes the form  
%
\begin{eqnarray}
\label{ansatz2}
\rho_+&=&1-(126-126\gamma_{+}+a_{10}+5a_{11})s^5
\nonumber\\
&+&(420-420\gamma_++5a_{10}+24a_{11})s^6
\nonumber\\
&-&(540-540\gamma_++10a_{10}+45a_{11})s^7
\nonumber\\
&+&(315-315\gamma_++10a_{10}+40a_{11})s^8
\nonumber\\
&-&(70-70\gamma_++5a_{10}+15a_{11})s^9
\nonumber\\
&+&a_{10}s^{10}+a_{11}s^{11}.
\end{eqnarray}
To minimize the perturbation energy in Eq. (\ref{perturbationenergy}), three free parameters are introduced, 
\begin{eqnarray}
\label{ansatz3}
\rho_+&=&1-(126-126\gamma_++c_{10}+5c_{11}+15c_{12})s^5
\nonumber\\
&+&(420-420\gamma_++5c_{10}+24c_{11}+70c_{12})s^6
\nonumber\\
&-&(540-540\gamma_++10c_{10}+45c_{11}+126c_{12})s^7
\nonumber\\
&+&(315-315\gamma_++10c_{10}+40c_{11}+105c_{12})s^8
\nonumber\\
&-&(70-70\gamma_++5c_{10}+15c_{11}+35c_{12})s^9
\nonumber\\
&+&c_{10}s^{10}+c_{11}s^{11}+c_{12}s^{12}.
\end{eqnarray}
\end{appendix}

\end{document}